\documentclass[journal=cmatex,manuscript=article]{achemso}

\usepackage{cleveref}
\usepackage{caption}
\usepackage{xkeyval}
\usepackage{geometry}
\usepackage{setspace}
\usepackage{achemso}
\usepackage[sort&compress]{natbib}
\SectionNumbersOn

\title{Polymorphic nanowires of magnetic shape memory MnAs} 

\author{C. Echeverr\'ia-Arrondo}
\email{carlos.echeverria@unavarra.es}
\affiliation{Departamento de F\'isica, Universidad P\'ublica de Navarra, E-31006, Pamplona, Spain}
\alsoaffiliation{Donostia International Physics Center (DIPC),  E-20018, San Sebasti\'an/Donostia, Spain}
\author{J. P\'erez-Conde}
\affiliation{Departamento de F\'isica, Universidad P\'ublica de Navarra, E-31006, Pamplona, Spain}
\author{A. Ayuela}
\affiliation{Centro de F\'isica de Materiales CFM-MPC Centro Mixto CSIC-UPV/EHU,  Departamento de
F\'isica de Materiales, E-20018, San Sebasti\'an/Donostia, Spain}
\alsoaffiliation{Donostia International Physics Center (DIPC),  E-20018, San Sebasti\'an/Donostia, Spain}

\begin{document}

\begin{abstract}

For nanostructured materials, strain is of fundamental
importance in stabilizing a specific crystallographic phase, modifying electronic properties, and in consequence their magnetism when
it applies. Here we describe a magnetic shape memory alloy in which nanostructure confinement strain influences the crystallographic phase and the electronic and
magnetic properties of the resulting nanowires. We use first-principles calculations on shape memory MnAs nanostructures to study  the influence of strain
on crystal phases, which arises from surfaces. We show that MnAs nanowires down to two nanometers can be stable in a new crystal phase different from bulk hexagonal and induced by one-dimensionality. The changes between structures through use of strain require the existence of twin domains. Our analysis suggests that the strain-induced structural transition, demonstrated here in MnAs compounds, could be extended to other (magnetic) shape memory nanowire systems for applications ranging from mechanical to magneto-electronic devices.

\end{abstract}
\maketitle



Nanocrystalline structures such as quantum wires  are attractive because of their peculiar properties arising from low dimensionality and quantum confinement which make them interesting for new nanodevice design \cite{huang,xia,morales,yu,holmes}. These nanocontacts, when synthesized from semiconductors such as Si,\cite{silicon} CdS\cite{cds} of group II-VI, and GaN\cite{gan} of group III-V, are promising in nanoelectronics. Ferromagnetic nanowires fabricated with the traditional materials Co, Fe and Ni\cite{sorop,thurn,whitney} and some alloys such as MnAs\cite{toyli} hold promise for ultra-high density recording devices \cite{simonds,sorop} in nanomagnetism. In the field between semiconductor electronics and magnetism, 1D nanostructures involving magnetic materials add interesting functionalities promising for spintronics. Wires fully made of diluted magnetic semiconductor GaAs:Mn, typical spintronic material, have already  been grown.\cite{sadowski} When looking at epitaxial semiconductor-magnet interfaces, semiconductor wires have attached magnetic dots of shape memory alloy MnAs \cite{ramlan} or are fully covered with MnAs wires around cores of III-V semiconductor GaAs.\cite{hilse}  Ferromagnetic nanowires made of MnAs are also fabricated on GaAs surfaces using lithographic\cite{toyli} and epitaxial\cite{engel} techniques. In all cases, the resulting MnAs/GaAs heterojunctions are clean, atomically sharp, and thermodynamically stable. \cite{tanaka, rungger} In these hybrid systems, the magnetic shape memory alloys such as MnAs are easily grown on semiconductor surfaces due to commensurability at the interface. The shape memory nanostructures made of MnAs can be highly strained because of nanostructure confinement, so strains will induce changes in other properties as it has already  been found for semiconductor nanocrystals.\cite{newton} Thus, the theoretical study of nanowires made of magnetic shape memory alloys is necessary to understand such low-dimensional quantum structures.

Magnetic shape memory alloys grow easily commensurate on semiconductor surfaces due to the softness of their phase transitions, which involve several crystal structures. This softness comes from the reversible transformation between a high-temperature phase called austenite and a low-temperature phase called martensite.\cite{handley} Recent experimental work has shown that the two  phases involved in the non-magnetic shape memory ternary alloy CuAlNi are more stable in nanoscale pillars than in the bulk.\cite{juan} If we focus on the magnetic shape memory alloy MnAs, the high-temperature and less-ordered orthorhombic phase changes at low temperatures to an hexagonal NiAs-type ferromagnetic phase.\cite{bean} Both structures are related by a shear of the atomic lattice together with relaxation of the unit cell atoms. Above 318 K the hexagonal symmetry transforms into orthorhombic MnP type and the ferromagnetic order of the Mn spins changes to antiferromagnetic, with  spin-up and spin-down Mn planes randomly distributed.\cite{occam} This hexagonal-to-orthorhombic transformation yields shape-memory effect in MnAs.\cite{bean} Although zinc-blende MnAs is unstable in the bulk, it has been successfully grown in the form of quantum dots\cite{fujimori} and thin films.\cite{film} Within the current investigation on magnetic nanowires made of shape memory MnAs, we address the important question concerning the stability of crystal phases under confinement, and how the crystal structure changes their electronic and magnetic properties.

In this work we investigate polymorphic shape memory nanowires made of MnAs within density functional theory. In particular, we study the geometrical stability of hexagonal NiAs-type and zinc-blende crystal structures. Unexpectedly, the most stable phase is neither the hexagonal nor the zinc blende, but a different structure induced by the low dimensionality of MnAs nanowires. This ground-state structure is characterized by 8 nearest neighbors around central Mn atoms and it is call hereafter ``8 index''. The fact that the crystal ground state is neither hexagonal nor zinc blende is surprising, and shows how strain plays an important role in stabilizing 1D crystal structures. To understand this confinement-induced behavior, we  analyze the strain of ferromagnetic MnAs nanowires through their associated winding numbers.\cite{mermin} For instance, we show that MnAs nanowires can not be \textit{continuously} deformed from the 8-index phase into zinc blende due to different winding numbers. Thus, the formation of either dislocations or twins in the zinc-blende phase under strain justifies the different energetic order. Additionally, we show that ferromagnetic MnAs nanowires are metallic in the 8-index phase with large carriers velocities at the Fermi energy, while hexagonal and cubic phases are half metallic.


\label{tc}

\begin{figure}[b]
\centering
\includegraphics[scale=0.7]{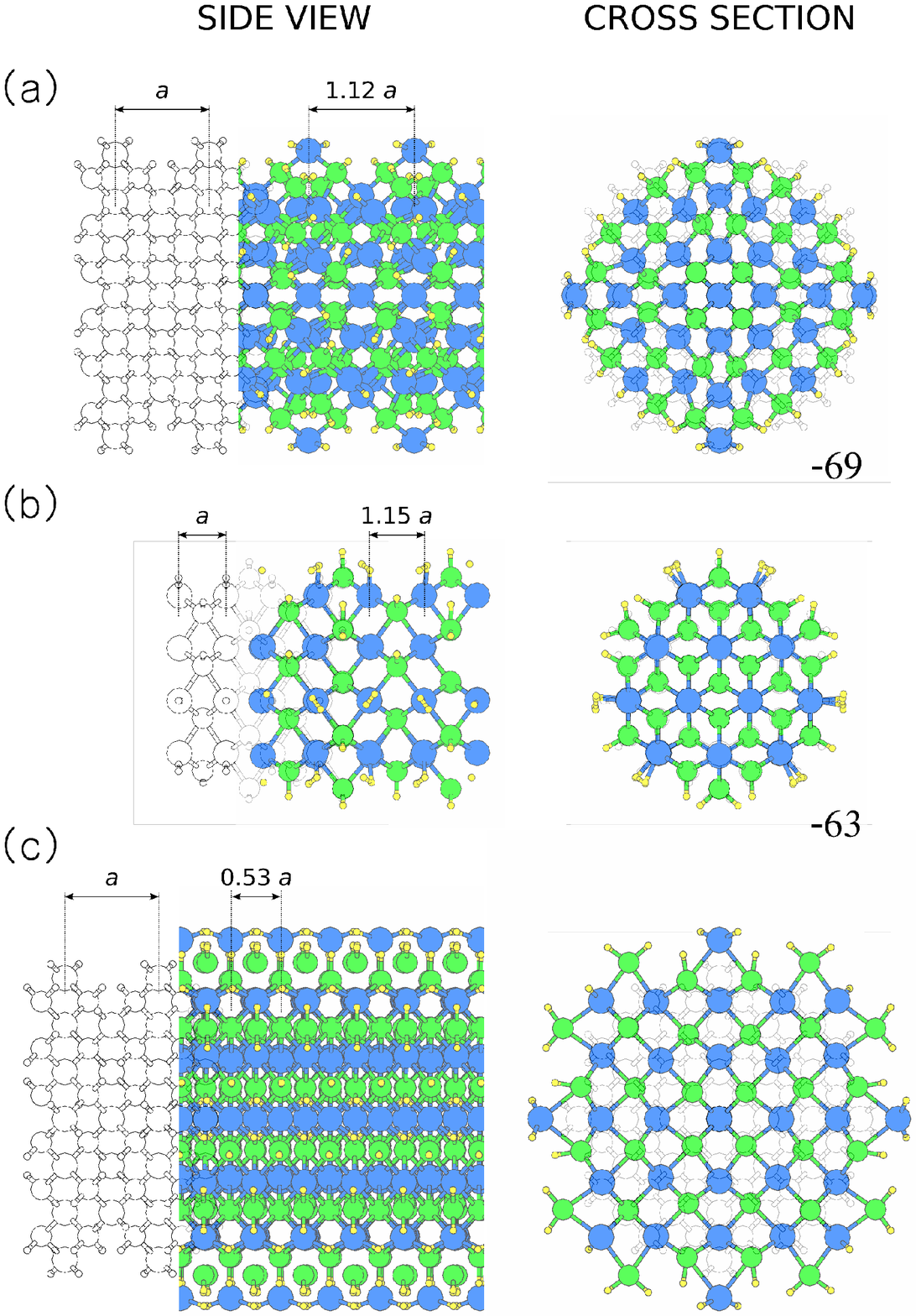}
\caption{\label{fig:fig1}(Color online) Side and cross-section views of polymorphic MnAs nanowires with (a) cubic zinc-blende, (b) hexagonal NiAs-type, and (c) the new crystal symmetry. The Mn and As atoms are drawn with large spheres in dark and light gray (blue and green), respectively; and pseudohydrogen atoms with small circles (yellow). The nanowires are at the same scale. For comparison, the stable crystal phases are depicted as in the bulk with empty spheres. The differences of cohesive energies per atom between wires and the 8-index found phase are given in meV below the plotted cubic and hexagonal geometries. The difference values show that although hexagonal wires are more stable than zinc-blende ones, these two phases are almost degenerate.}
\end{figure}

We investigate the threefold polymorphism of ferromagnetic MnAs nanowires by relaxing two bulk-like input geometries, hexagonal and zinc blende, as depicted in Fig.~\plainref{fig:fig1}, where small pseudohydrogens\cite{biblia} are included in order to avoid surface effects and mimic the surrounding environment. In addition, the wires are studied in the ferromagnetic configuration of Mn spins since the antiferromagnetic ordering is less stable for the three shown crystal phases. See below section concerning method and computations for further details.

%

First we look at zinc blende nanowires when streching or compressing with different c/a ratios. Our output zinc-blende nanowires are drawn in Fig.~\plainref{fig:fig1}(a). During relaxation, the Mn-As bonds stretch from 2.46~{\AA} to $2.49-2.61$~{\AA} and the wire axis expands by 12\%. The shape of the cross section changes from a circle  to a rhombus. This output shape calls for a more detailed analysis of zinc-blende wires, as given later. The input and output hexagonal structures are plotted in Fig.~\plainref{fig:fig1}(b) with the $\langle0001\rangle$ axis parallel to the wire axis ($c$-plane orientation). We choose cross sections terminated with As edges since they are more stable than those with Mn.\cite{prb-79-195420} During relaxation, the Mn-As bonds lenghts increase from 2.57~{\AA} to $2.64-2.77$~{\AA}. This large strain (3$-$8\%) compares to lattice mismatches measured at the MnAs/GaAs interface ($<30\%$\cite{mismatch}). Due to relaxation strain, the hexagonal wires expand along the axis by 15\%. The 8-index phase is obtained by structural relaxation of a compressed zinc-blende wire along the axis with a small c/a ratio. This output geometry is drawn in Fig.~\plainref{fig:fig1}(c) together with its bulk-like zinc-blende counterpart. In the relaxed wires, the Mn-As bonds range from 2.68~{\AA} to 3.13~{\AA} in length. As compared with the bulk-like zinc-blende structure, the 8-index phase is contracted along the axis down to 0.53$a$. These relaxed wires are characterized by 8 nearest neighbors around the Mn atoms in the axis. For the As and Mn atoms off-center, the coordination index is either 5 or 6 depending on their  positions. 

The cohesive energy per atom in the unit cell is defined as $-(E_0 - \sum_{i=1}^N E_{\rm{at}}^i)/N$, where $E_0$ is the total energy and $E_{\rm{at}}^i$ is the energy of the $i$-th free atom. The calculated values are 2.698, 2.635, and 2.629~eV for 8-index, hexagonal, and cubic phases, respectively. Therefore, the MnAs nanowires are more stable with the new crystal geometry than with either the hexagonal or zinc-blende one, in the same order. The fact that a new phase different from bulk hexagonal is more stable shows the important role of strain induced by 1D nanostructures. 

\begin{figure}[!t]
\centering
\includegraphics[scale=0.8]{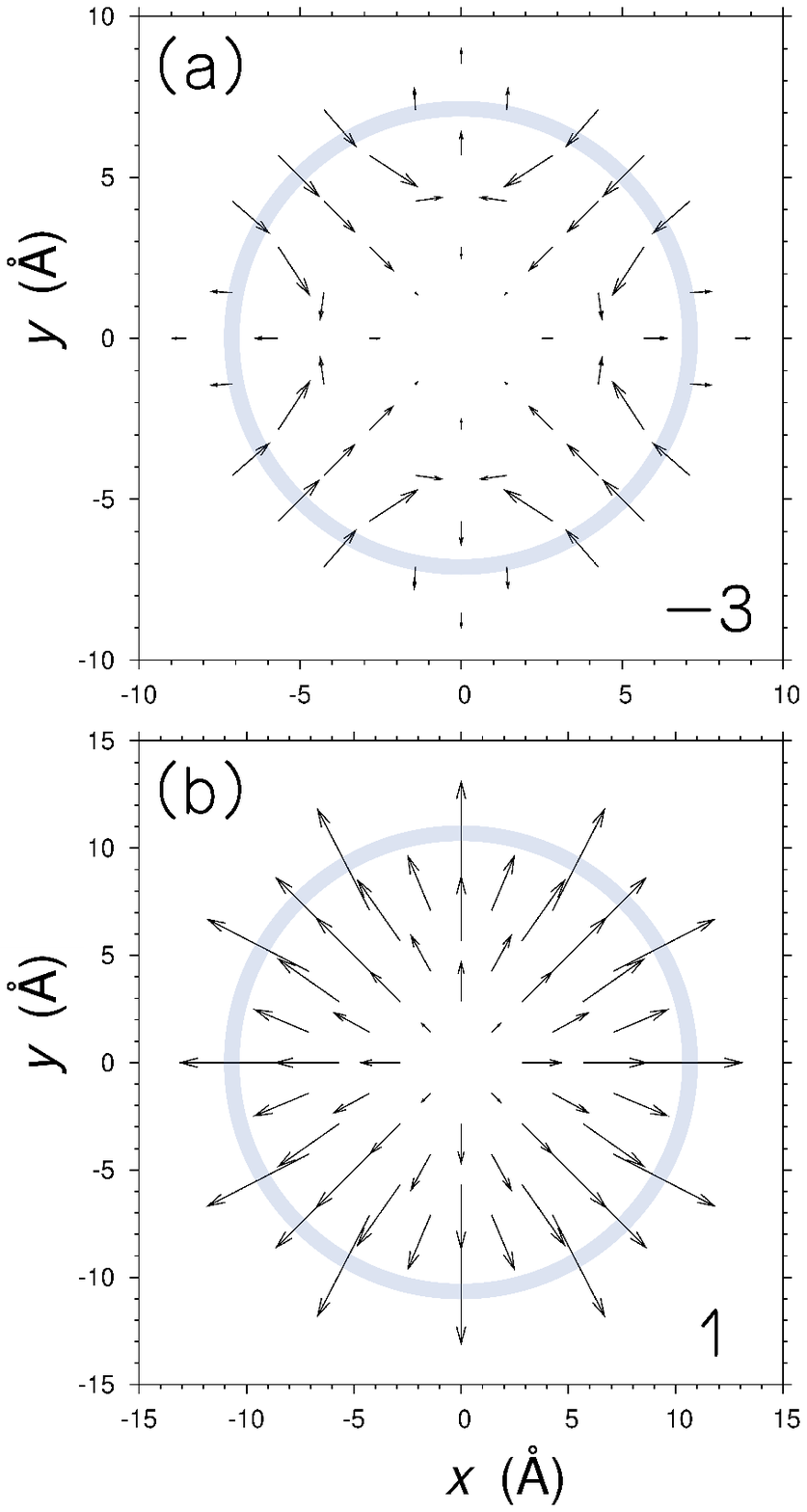}
\caption{\label{fig:fig2}(Color online) Strain fields in the cross sections of (a) cubic and (b)  8-index nanowires. The strain vectors are multiplied by 2 for the sake of clarity. The numbers in the down right corners are the winding numbers around the nanowire surfaces. These numbers are the times that the strain vectors on the marked circumferences, when looking counterclockwise, rotate. Counterclockwise (clockwise) rotations give positive (negative) winding numbers. Different winding numbers suggest the presence of either dislocations of twins in the zinc-blende nanowires.}
\end{figure}

Let us further investigate the nature of polymorphism in MnAs nanowires  by looking at their strain fields and associated winding numbers. The structural relaxation of the nanowires determine strain vectors that form strain fields in their cross sections. We focus on the 8-index and zinc-blende nanowires, as drawn in Fig.~\plainref{fig:fig2}. The strain field corresponding to the ground state shows an homogeneous expansion that can be explained because of the compressed input geometry. However, the expansion for the zinc-blende phase is clearly inhomogeneous and it must be analyzed in more detail.

These strain fields are now studied within homotopy theory.\cite{mermin} This theory provides the formalism for the precise description of a large variety of ordered systems through a vectorial function. For strain fields, the values of this function are the strain vectors lying on circumferences centered in nanowire axis. The vectorial function is here not continuous but discretized at every atom.  When looking counterclockwise at these strain vectors, the times that they turn around is known as the winding number of the strain field. This number is positive (negative) for counterclockwise (clockwise) rotations. For 8-index and cubic geometries, the winding numbers are 1 and -3, respectively. From homotopy theory we know that crystal structures with distinct winding numbers are not homotopic, that is, they can not be \textit{continuously} deformed into one another. Since 8-index and cubic structures are not homotopic, the phase transition between them suggests the formation of dislocations, such as twins, in the zinc-blende wires. 

Thus, with a larger number of dislocations, the energy of cubic structure being less stable than the 8-index one, we are lead to the idea of an additional dislocation energy in defective structures. This energy is related to the number of stacking faults in the nanowires. Given these dislocations, we can return to the cohesive energy differences of eV per atom between 8-index and other geometries. We can qualitatively understand this as follows. Assumed that the dislocation energy is $\xi$ ($~$ 20 mev about tenths of meV for pure metals\cite{cu})  and that the number of dislocations is, say, $n$ ($~ 3$ for zinc-blende wires), then the cohesive energy per atom is of the order of $n\xi\simeq60$~meV. This value is very close to our calculated cohesive differences, as shown in Fig.~\plainref{fig:fig1}.


We now study the ferromagnetism of MnAs nanowires in more detail. The calculated total magnetic moments per Mn atom are 4.00$\mu_B$, 4.38$\mu_B$, and 4.59$\mu_B$ in the cubic, hexagonal, and 8-index nanowires. For comparison, we also calculate the total magnetic moments in the bulk, 4.00$\mu_B$ for zinc-blende MnAs and $\sim$3.89$\mu_B$ for hexagonal MnAs. This latter value is close to the experimental one for the hexagonal phase, $\sim$3.4$\mu_B$;\cite{bean,goodenough,menyuk} we remember that bulk zinc-blende MnAs is unstable. The previous magnetic moments 4.00$\mu_B$ and $\sim$3.89$\mu_B$ show the same decreasing trend than 3.75$\mu_B$ and 3.09$\mu_B$ reported for bulk  MnAs with zinc-blende and hexagonal structures,\cite{continenza} respectively. In comparison with previous structural differences in magnetic moments,  the position dependence is smaller, in the order of 0.1$\mu_B$, and it can neglected. 


The calculated total magnetic moment of 4.00$\mu_B$ per Mn atom in the zinc-blende nanowires can be explained from (i) the large charge transfer of near three electrons from the Mn cations to their neighbor As anions, and (ii) the strong $p$(As)$-d$(Mn) hybridization.\cite{sanvito} These two effects are in turn strongly dependent on the Mn-As bond length which is  $\sim$2.55~{\AA} in the zinc-blende phase, close to 2.46~{\AA} in the bulk.\cite{freeman-2} Looking at coordination numbers the total magnetic moment in the hexagonal phase  should be smaller than that in the zinc-blende structure. However, the previous 4.38$\mu_B$ value for the hexagonal wires is larger than 4.00$\mu_B$ for the zinc-blende ones and also larger than $\sim3.89\mu_B$ in the bulk. Both the charge transfer from Mn atoms and the $p-d$ hybridization are reduced because of the longer Mn-As bond length, $\sim$2.70~{\AA}. Similarly, the 4.59$\mu_B$ total magnetic moment for the 8-index phase is related to the even longer Mn-As bond distance, $\sim$2.90~{\AA}. 

\begin{figure}[!t]
\centering
\includegraphics[scale=0.55]{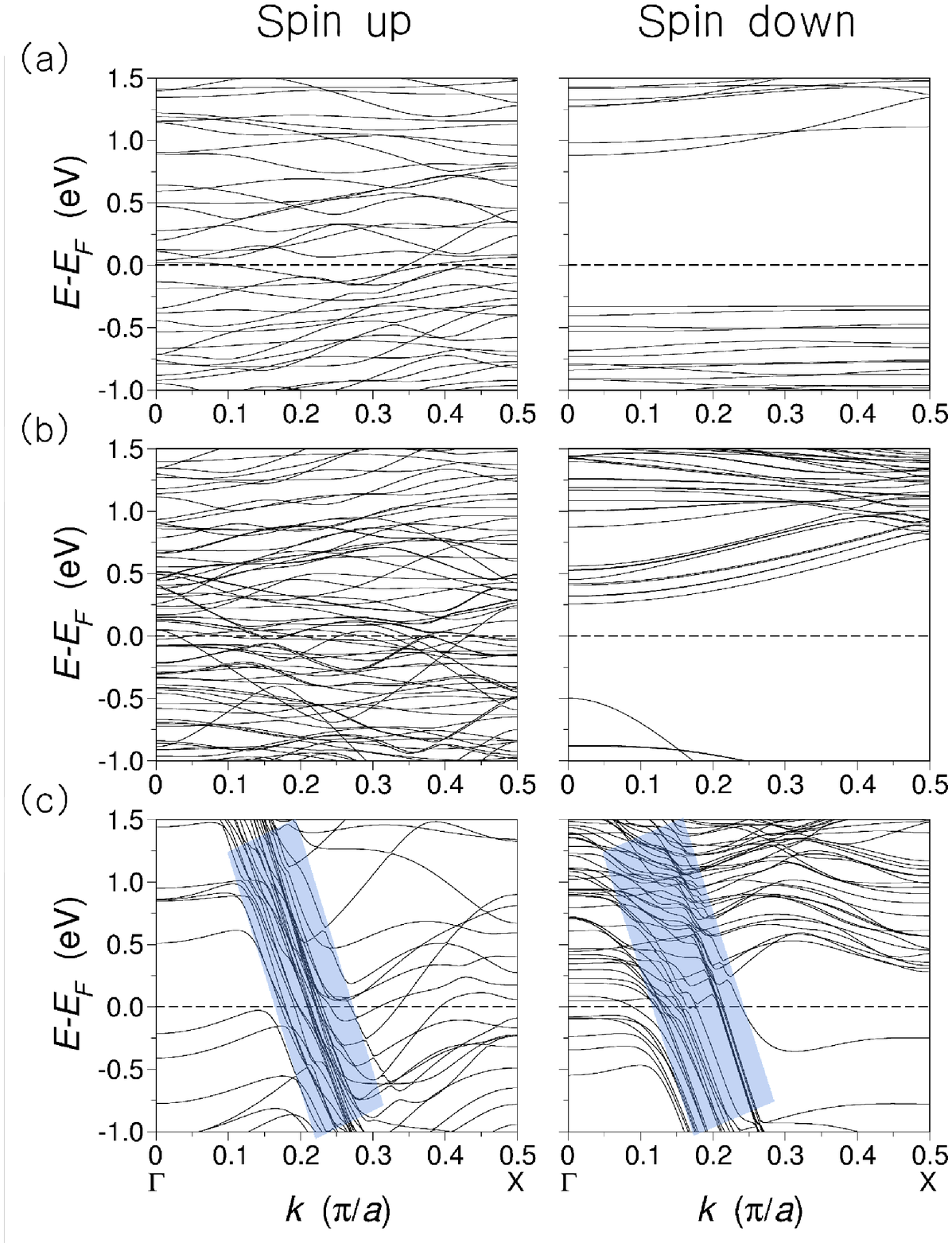}
\caption{\label{fig:fig3}(Color online) Spin-polarized electronic bands calculated for (a) cubic, (b) hexagonal, and (c) 8-index MnAs nanowires. The shadowed areas indicate the slopes for the 8-index bands crossing the Fermi level. Note that carriers velocities are thus larger for 8-index wires than for either hexagonal or cubic ones.}
\end{figure}

Next we are interested in the electronic properties of MnAs nanowires. The electronic bands along the wire direction are drawn in Fig.~\plainref{fig:fig3} as a function of the electron wave vector between $\Gamma$ and X points. In the $E(k)$ curves, we see that the hexagonal and zinc-blende nanowires behave as half-metals due to a gap in the spin-down channel, and that the 8-index ones behave as metals. We note that zinc-blende MnAs is also half-metallic in thin films.\cite{film} From the slope of the dispersion curves we also estimate the velocities of the carriers close to the Fermi energy. The largest slopes for both spin components correspond to the 8-index wires, as shadowed in Fig.~\plainref{fig:fig3}(c). The carriers velocities are thus larger for this new found structure than for the hexagonal and cubic ones. By tuning the crystal geometry with strain we can change the polytypism and twin planes inside nanowires. This level of control could lead to different transport properties and novel electronic behavior, recently found in semiconductor wires.\cite{caroff,gudiksen}


In summary, we have investigated the nature of polymorphism in ferromagnetic nanowires made of shape memory MnAs within density functional theory. The experimental phases for bulk MnAs and nanostructures are hexagonal NiAs type and zinc blende. However, the MnAs nanowires are more stable in the so-called "8-index"  phase induced by strain in one dimension. The 8-index geometry has been further investigated by comparing its strain field with that of the zinc-blende wires. The strain fields are formed by the strain vectors arising from structural relaxation of the atomic positions. Within homotopy theory,\cite{mermin} we have looked counterclockwise at the strain vectors lying on the most external circunferences centered in the wires and counted the times that these vectors rotate. For a strain field, this number is its associated winding number. The fact that 8-index and zinc-blende wires have distinct winding numbers suggests the formation in the latter of either twins or stacking faults. Moreover, the calculated electronic bands show that zinc-blende and hexagonal phases are half-metallic, while the strain-induced most stable crystal structure is metallic with also large velocities for the carriers. Our results thus bring into contact the threefold polymorphism of MnAs nanowires with their electronic and magnetic properties, and they broaden the applicability of shape memory MnAs in nanomagnetism and spin-based nanoelectronics.

\section{METHOD AND COMPUTATIONAL DETAILS}
\label{tc}

We investigate the polymorphism of magnetic nanowires made of shape memory MnAs within density functional theory. We calculate their geometrical, electronic, and magnetic properties with the projector augmented-wave method, as implemented in VASP (Vienna Ab-initio Simulation Package).\cite{kresse1,kresse2,kresse3,kresse4} We use the GGA+$U$ exchange-correlation approach\cite{pbe} which models the strong interactions between the Mn 3$d$ electrons through the $U$ Coulomb and $J$ exchange parameters, $U=4$~eV and $J=0.8$~eV.\cite{park}

We study MnAs nanowires in three different stable phases, namely 8-index, hexagonal, and cubic zinc blende. Their respective cross sections measure about 22, 13, and 18~{\AA} in diameter, as given in Fig.~\plainref{fig:fig1}, where we also show that surface dangling bonds are passivated with pseudohydrogens.\cite{biblia} Through passivation we reproduce the wires surrounding material and we avoid the appearence of surface states in the gap  region. Independently of the crystal geometry, the saturated dangling bonds are tetrahedrally arranged in space in order to reproduce the four-fold coordination of surrounding semiconductors such as coats, matrices, and substrates like GaAs\cite{kas}.

We fix the cut-off energy in the plane-wave basis set at 400~eV in order to converge the total energies calculated for bulk MnAs with hexagonal and cubic structures within meVs. For the hexagonal case, we use the lattice constants $a=3.7$~{\AA} and $c=5.7$~{\AA};\cite{schippan} for the zinc-blende case, $a=5.68$~{\AA}.\cite{freeman-2} The atomic positions in the ferromagnetic MnAs nanowires are relaxed until the forces on the atoms are smaller than 0.02 eV/{\AA}. We take six \textit{k} points in the $\Gamma$-X region of the first Brillouin zone to calculate relaxations, total energies, and charge densities. The number of \textit{k} points is increased up to fifty in order to plot the electronic bands.

Apart from ferromagnetic (FM) nanowires, we also investigate to a first approximation the antiferromagnetic (AFM) configuration of Mn spins. This magnetic ordering arises from up and down Mn planes. We define the exchange energy as $E_0(\rm{AFM})$-$E_0(\rm{FM})$, where $E_0$ is the total energy. For 8-index, hexagonal, and cubic nanowires, the calculated exchange values are 16, 156, and 152~meV/Mn atom, respectively. The ferromagnetic Mn spins are thus more stable than the antiferromagnetic, and the values for the hexagonal and cubic cases are almost degenerate, as it occurs for their cohesive energies.


\newpage
\begin{thebibliography}{46}
\expandafter\ifx\csname natexlab\endcsname\relax\def\natexlab#1{#1}\fi
\expandafter\ifx\csname bibnamefont\endcsname\relax
  \def\bibnamefont#1{#1}\fi
\expandafter\ifx\csname bibfnamefont\endcsname\relax
  \def\bibfnamefont#1{#1}\fi
\expandafter\ifx\csname citenamefont\endcsname\relax
  \def\citenamefont#1{#1}\fi
\expandafter\ifx\csname url\endcsname\relax
  \def\url#1{\texttt{#1}}\fi
\expandafter\ifx\csname urlprefix\endcsname\relax\def\urlprefix{URL }\fi
\providecommand{\bibinfo}[2]{#2}
\providecommand{\eprint}[2][]{\url{#2}}

\bibitem[{\citenamefont{Huang et~al.}(2001)\citenamefont{Huang, Mao, Feick,
  Yan, Wu, Kind, Weber, Russo, and Yang}}]{huang}
\bibinfo{author}{\bibfnamefont{M.~H.} \bibnamefont{Huang}},
  \bibinfo{author}{\bibfnamefont{S.}~\bibnamefont{Mao}},
  \bibinfo{author}{\bibfnamefont{H.}~\bibnamefont{Feick}},
  \bibinfo{author}{\bibfnamefont{H.}~\bibnamefont{Yan}},
  \bibinfo{author}{\bibfnamefont{Y.}~\bibnamefont{Wu}},
  \bibinfo{author}{\bibfnamefont{H.}~\bibnamefont{Kind}},
  \bibinfo{author}{\bibfnamefont{E.}~\bibnamefont{Weber}},
  \bibinfo{author}{\bibfnamefont{R.}~\bibnamefont{Russo}}, \bibnamefont{and}
  \bibinfo{author}{\bibfnamefont{P.}~\bibnamefont{Yang}},
  \bibinfo{journal}{\textit{Science}} \textbf{2001} \textit{\bibinfo{volume}{292}},
  \bibinfo{pages}{1897-1899} 

\bibitem[{\citenamefont{Xia et~al.}(2003)\citenamefont{Xia, Yang, Sun, Wu,
  Mayers, Gates, Yin, Kim, and Yan}}]{xia}
\bibinfo{author}{\bibfnamefont{Y.}~\bibnamefont{Xia}},
  \bibinfo{author}{\bibfnamefont{P.}~\bibnamefont{Yang}},
  \bibinfo{author}{\bibfnamefont{Y.}~\bibnamefont{Sun}},
  \bibinfo{author}{\bibfnamefont{Y.}~\bibnamefont{Wu}},
  \bibinfo{author}{\bibfnamefont{B.}~\bibnamefont{Mayers}},
  \bibinfo{author}{\bibfnamefont{B.}~\bibnamefont{Gates}},
  \bibinfo{author}{\bibfnamefont{Y.}~\bibnamefont{Yin}},
  \bibinfo{author}{\bibfnamefont{F.}~\bibnamefont{Kim}}, \bibnamefont{and}
  \bibinfo{author}{\bibfnamefont{H.}~\bibnamefont{Yan}}, \bibinfo{journal}{\textit{Adv.
  Mater.}} \textbf{2003}, \textit{\bibinfo{volume}{15}}, \bibinfo{pages}{353-389}
  

\bibitem[{\citenamefont{Morales and Lieber}(1998)}]{morales}
\bibinfo{author}{\bibfnamefont{A.~M.} \bibnamefont{Morales}} \bibnamefont{and}
  \bibinfo{author}{\bibfnamefont{C.~M.} \bibnamefont{Lieber}},
  \bibinfo{journal}{\textit{Science}} \textbf{1998}, \textit{\bibinfo{volume}{279}},
  \bibinfo{pages}{208-211}

\bibitem[{\citenamefont{Yu et~al.}(2003)\citenamefont{Yu, Li, Loomis, Wang, and
  Buhro}}]{yu}
\bibinfo{author}{\bibfnamefont{H.}~\bibnamefont{Yu}},
  \bibinfo{author}{\bibfnamefont{J.}~\bibnamefont{Li}},
  \bibinfo{author}{\bibfnamefont{R.~A.} \bibnamefont{Loomis}},
  \bibinfo{author}{\bibfnamefont{L.~W.} \bibnamefont{Wang}}, \bibnamefont{and}
  \bibinfo{author}{\bibfnamefont{W.~E.} \bibnamefont{Buhro}},
  \bibinfo{journal}{\textit{Nature Mat.}} \textbf{2003}, \textit{\bibinfo{volume}{2}},
  \bibinfo{pages}{517-520}.

\bibitem[{\citenamefont{Holmes et~al.}(2000)\citenamefont{Holmes, Johnston, C.,
  and Korgel}}]{holmes}
\bibinfo{author}{\bibfnamefont{J.~D.} \bibnamefont{Holmes}},
  \bibinfo{author}{\bibfnamefont{K.~P.} \bibnamefont{Johnston}},
  \bibinfo{author}{\bibfnamefont{D.~R.} \bibnamefont{C.}}, \bibnamefont{and}
  \bibinfo{author}{\bibfnamefont{B.~A.} \bibnamefont{Korgel}},
  \bibinfo{journal}{\textit{Science}} \textbf{2000}, \textit{\bibinfo{volume}{287}},
  \bibinfo{pages}{1471-1473}.

\bibitem[{\citenamefont{Cui et~al.}(2001)\citenamefont{Cui, Lauhon, Gudiksen,
  Wang, and Lieber}}]{silicon}
\bibinfo{author}{\bibfnamefont{Y.}~\bibnamefont{Cui}},
  \bibinfo{author}{\bibfnamefont{L.~J.} \bibnamefont{Lauhon}},
  \bibinfo{author}{\bibfnamefont{M.~S.} \bibnamefont{Gudiksen}},
  \bibinfo{author}{\bibfnamefont{J.}~\bibnamefont{Wang}}, \bibnamefont{and}
  \bibinfo{author}{\bibfnamefont{C.~M.} \bibnamefont{Lieber}},
  \bibinfo{journal}{\textit{Appl. Phys. Lett.}} \textbf{2001}, \textit{\bibinfo{volume}{78}},
  \bibinfo{pages}{2214-2216}.

\bibitem[{\citenamefont{Barrelet et~al.}(2003)\citenamefont{Barrelet, Wu, Bell,
  and Lieber}}]{cds}
\bibinfo{author}{\bibfnamefont{C.~J.} \bibnamefont{Barrelet}},
  \bibinfo{author}{\bibfnamefont{Y.}~\bibnamefont{Wu}},
  \bibinfo{author}{\bibfnamefont{D.~C.} \bibnamefont{Bell}}, \bibnamefont{and}
  \bibinfo{author}{\bibfnamefont{C.~M.} \bibnamefont{Lieber}},
  \bibinfo{journal}{\textit{J. Am. Chem. Soc.}} \textbf{2003}, \textit{125}, \bibinfo{pages}{11498-11499}


\bibitem[{\citenamefont{Huang et~al.}(2002)\citenamefont{Huang, Duan, Cui, and
  Lieber}}]{gan}
\bibinfo{author}{\bibfnamefont{Y.}~\bibnamefont{Huang}},
  \bibinfo{author}{\bibfnamefont{X.}~\bibnamefont{Duan}},
  \bibinfo{author}{\bibfnamefont{Y.}~\bibnamefont{Cui}}, \bibnamefont{and}
  \bibinfo{author}{\bibfnamefont{C.~M.} \bibnamefont{Lieber}},
  \bibinfo{journal}{\textit{Nano Lett.}} \textbf{2002}, \textit{\bibinfo{volume}{2}},
  \bibinfo{pages}{101-104}.

\bibitem[{\citenamefont{Sorop et~al.}(2003)\citenamefont{Sorop, Untiedt, Luis,
  Kr{\"o}ll, Rasa, and {de Jongh}}}]{sorop}
\bibinfo{author}{\bibfnamefont{T.~G.} \bibnamefont{Sorop}},
  \bibinfo{author}{\bibfnamefont{C.}~\bibnamefont{Untiedt}},
  \bibinfo{author}{\bibfnamefont{F.}~\bibnamefont{Luis}},
  \bibinfo{author}{\bibfnamefont{M.}~\bibnamefont{Kr{\"o}ll}},
  \bibinfo{author}{\bibfnamefont{M.}~\bibnamefont{Rasa}}, \bibnamefont{and}
  \bibinfo{author}{\bibfnamefont{L.~J.} \bibnamefont{{de Jongh}}},
  \bibinfo{journal}{\textit{Phys. Rev. B}} \textbf{2003}, \textit{\bibinfo{volume}{67}},
  \bibinfo{pages}{014402-014409}.

\bibitem[{\citenamefont{Thurn-Albrecht
  et~al.}(2000)\citenamefont{Thurn-Albrecht, Schotter, K{\"a}stle, Emley,
  Shibauchi, Krusin-Elbaum, Guarini, Black, Tuominen, and Russell}}]{thurn}
\bibinfo{author}{\bibfnamefont{T.}~\bibnamefont{Thurn-Albrecht}},
  \bibinfo{author}{\bibfnamefont{J.}~\bibnamefont{Schotter}},
  \bibinfo{author}{\bibfnamefont{G.~A.} \bibnamefont{K{\"a}stle}},
  \bibinfo{author}{\bibfnamefont{N.}~\bibnamefont{Emley}},
  \bibinfo{author}{\bibfnamefont{T.}~\bibnamefont{Shibauchi}},
  \bibinfo{author}{\bibfnamefont{L.}~\bibnamefont{Krusin-Elbaum}},
  \bibinfo{author}{\bibfnamefont{K.}~\bibnamefont{Guarini}},
  \bibinfo{author}{\bibfnamefont{C.~T.} \bibnamefont{Black}},
  \bibinfo{author}{\bibfnamefont{M.~T.} \bibnamefont{Tuominen}},
  \bibnamefont{and} \bibinfo{author}{\bibfnamefont{T.~P.}
  \bibnamefont{Russell}}, \bibinfo{journal}{\textit{Science}} \textbf{2000},
  \textit{\bibinfo{volume}{290}}, \bibinfo{pages}{2126-2129}.

\bibitem[{\citenamefont{Whitney et~al.}(1993)\citenamefont{Whitney, Jiang,
  Searson, and Chien}}]{whitney}
\bibinfo{author}{\bibfnamefont{T.~M.} \bibnamefont{Whitney}},
  \bibinfo{author}{\bibfnamefont{J.~S.} \bibnamefont{Jiang}},
  \bibinfo{author}{\bibfnamefont{P.~C.} \bibnamefont{Searson}},
  \bibnamefont{and} \bibinfo{author}{\bibfnamefont{C.~L.} \bibnamefont{Chien}},
  \bibinfo{journal}{\textit{Science}} \textbf{1993}, \textit{\bibinfo{volume}{\textit{261}}},
  \bibinfo{pages}{1316-1319}.

\bibitem[{\citenamefont{Toyli}(2008)}]{toyli}
\bibinfo{author}{\bibfnamefont{D.}~\bibnamefont{Toyli}}, \bibinfo{journal}{\textit{J.
  Young Investigators}} \textbf{2008}, \textit{\bibinfo{volume}{16}},
  \urlprefix\url{http://www.jyi.org/research/re.php?id=964}.

\bibitem[{\citenamefont{Simonds}(1995)}]{simonds}
\bibinfo{author}{\bibfnamefont{J.~L.} \bibnamefont{Simonds}},
  \bibinfo{journal}{\textit{Phys. Today}} \textbf{1995}, \textit{\bibinfo{volume}{48}},
  \bibinfo{pages}{26-32}.

\bibitem[{\citenamefont{Sadowski et~al.}(2007)\citenamefont{Sadowski,
  Dlu\.zewski, Kret, Janik, Lusakowska, Kanski, Presz, Terki, Charar, and
  Tang}}]{sadowski}
\bibinfo{author}{\bibfnamefont{J.}~\bibnamefont{Sadowski}},
  \bibinfo{author}{\bibfnamefont{P.}~\bibnamefont{Dlu\.zewski}},
  \bibinfo{author}{\bibfnamefont{S.}~\bibnamefont{Kret}},
  \bibinfo{author}{\bibfnamefont{E.}~\bibnamefont{Janik}},
  \bibinfo{author}{\bibfnamefont{E.}~\bibnamefont{Lusakowska}},
  \bibinfo{author}{\bibfnamefont{J.}~\bibnamefont{Kanski}},
  \bibinfo{author}{\bibfnamefont{A.}~\bibnamefont{Presz}},
  \bibinfo{author}{\bibfnamefont{F.}~\bibnamefont{Terki}},
  \bibinfo{author}{\bibfnamefont{S.}~\bibnamefont{Charar}}, \bibnamefont{and}
  \bibinfo{author}{\bibfnamefont{D.}~\bibnamefont{Tang}},
  \bibinfo{journal}{\textit{Nano Lett.}} \textbf{2007}, \textit{\bibinfo{volume}{7}},
  \bibinfo{pages}{2724-2728}.

\bibitem[{\citenamefont{Ramlan et~al.}(2006)\citenamefont{Ramlan, May, Zheng,
  Allen, Wessels, and Lauhon}}]{ramlan}
\bibinfo{author}{\bibfnamefont{D.~G.} \bibnamefont{Ramlan}},
  \bibinfo{author}{\bibfnamefont{S.~J.} \bibnamefont{May}},
  \bibinfo{author}{\bibfnamefont{J.-G.} \bibnamefont{Zheng}},
  \bibinfo{author}{\bibfnamefont{J.~E.} \bibnamefont{Allen}},
  \bibinfo{author}{\bibfnamefont{B.~W.} \bibnamefont{Wessels}},
  \bibnamefont{and} \bibinfo{author}{\bibfnamefont{L.~J.}
  \bibnamefont{Lauhon}}, \bibinfo{journal}{\textit{Nano Lett.}} \textbf{2006},
  \textit{\bibinfo{volume}{6}}, \bibinfo{pages}{50-54}.

\bibitem[{\citenamefont{Hilse et~al.}(2009)\citenamefont{Hilse, Takagaki,
  Herfort, Ramsteiner, Herrmann, Breuer, Geelhaar, and Riechert}}]{hilse}
\bibinfo{author}{\bibfnamefont{M.}~\bibnamefont{Hilse}},
  \bibinfo{author}{\bibfnamefont{Y.}~\bibnamefont{Takagaki}},
  \bibinfo{author}{\bibfnamefont{J.}~\bibnamefont{Herfort}},
  \bibinfo{author}{\bibfnamefont{M.}~\bibnamefont{Ramsteiner}},
  \bibinfo{author}{\bibfnamefont{C.}~\bibnamefont{Herrmann}},
  \bibinfo{author}{\bibfnamefont{S.}~\bibnamefont{Breuer}},
  \bibinfo{author}{\bibfnamefont{L.}~\bibnamefont{Geelhaar}}, \bibnamefont{and}
  \bibinfo{author}{\bibfnamefont{H.}~\bibnamefont{Riechert}},
  \bibinfo{journal}{\textit{Appl. Phys. Lett.}} \textbf{2009}, \textit{\bibinfo{volume}{95}},
  \bibinfo{pages}{133126-133128}.

\bibitem[{\citenamefont{Engel-Herbert et~al.}(2006)\citenamefont{Engel-Herbert,
  Hesjedal, Mohanty, Schaadt, and Ploog}}]{engel}
\bibinfo{author}{\bibfnamefont{R.}~\bibnamefont{Engel-Herbert}},
  \bibinfo{author}{\bibfnamefont{T.}~\bibnamefont{Hesjedal}},
  \bibinfo{author}{\bibfnamefont{J.}~\bibnamefont{Mohanty}},
  \bibinfo{author}{\bibfnamefont{D.~M.} \bibnamefont{Schaadt}},
  \bibnamefont{and} \bibinfo{author}{\bibfnamefont{K.~H.} \bibnamefont{Ploog}},
  \bibinfo{journal}{\textit{Phys. Rev. B}} \textbf{2006}, \textit{\bibinfo{volume}{73}},
  \bibinfo{pages}{104441-104447}.

\bibitem[{\citenamefont{Tanaka}(2002)}]{tanaka}
\bibinfo{author}{\bibfnamefont{M.}~\bibnamefont{Tanaka}},
  \bibinfo{journal}{\textit{Semi. Sci. Technol.}} \textbf{2002}, \textit{\bibinfo{volume}{17}},
  \bibinfo{pages}{327-341}.

\bibitem[{\citenamefont{Rungger and Sanvito}(2006)}]{rungger}
\bibinfo{author}{\bibfnamefont{I.}~\bibnamefont{Rungger}} \bibnamefont{and}
  \bibinfo{author}{\bibfnamefont{S.}~\bibnamefont{Sanvito}},
  \bibinfo{journal}{\textit{Phys. Rev. B}} \textbf{2006}, \textit{\bibinfo{volume}{74}},
  \bibinfo{pages}{024429-024442}.

\bibitem[{\citenamefont{Newton et~al.}(2010)\citenamefont{Newton, Leake,
  Harder, and Robinson}}]{newton}
\bibinfo{author}{\bibfnamefont{M.~C.} \bibnamefont{Newton}},
  \bibinfo{author}{\bibfnamefont{S.~T.} \bibnamefont{Leake}},
  \bibinfo{author}{\bibfnamefont{R.}~\bibnamefont{Harder}}, \bibnamefont{and}
  \bibinfo{author}{\bibfnamefont{I.~K.} \bibnamefont{Robinson}},
  \bibinfo{journal}{\textit{Nat. Mat.}} \textbf{2010}, \textit{\bibinfo{volume}{9}},
  \bibinfo{pages}{120-124}.

\bibitem[{\citenamefont{Handley}(2000)}]{handley}
\bibinfo{author}{\bibfnamefont{R.~C.~O.} \bibnamefont{Handley}},
  \emph{\bibinfo{title}{Modern Magnetic Materials: Principles and
  Applications}} (\bibinfo{publisher}{John Wiley \& sons},
  \bibinfo{address}{New York}, \bibinfo{year}{2000}).

\bibitem[{\citenamefont{Juan et~al.}(2009)\citenamefont{Juan, N{\'o}, and
  Schuh}}]{juan}
\bibinfo{author}{\bibfnamefont{J.~S.} \bibnamefont{Juan}},
  \bibinfo{author}{\bibfnamefont{M.~L.} \bibnamefont{N{\'o}}},
  \bibnamefont{and} \bibinfo{author}{\bibfnamefont{C.~A.} \bibnamefont{Schuh}},
  \bibinfo{journal}{\textit{Nature Nanotech.}} \textbf{2009}, \textit{\bibinfo{volume}{4}},
  \bibinfo{pages}{415-419}.

\bibitem[{\citenamefont{Bean and Rodbell}(1962)}]{bean}
\bibinfo{author}{\bibfnamefont{C.~P.} \bibnamefont{Bean}} \bibnamefont{and}
  \bibinfo{author}{\bibfnamefont{D.~S.} \bibnamefont{Rodbell}},
  \bibinfo{journal}{\textit{Phys. Rev.}} \textbf{1962}, \textit{\bibinfo{volume}{126}},
  \bibinfo{pages}{104-115}.

\bibitem[{\citenamefont{Niranjan et~al.}(2004)\citenamefont{Niranjan, Sahu, and
  Kleinman}}]{occam}
\bibinfo{author}{\bibfnamefont{M.~K.} \bibnamefont{Niranjan}},
  \bibinfo{author}{\bibfnamefont{B.~R.} \bibnamefont{Sahu}}, \bibnamefont{and}
  \bibinfo{author}{\bibfnamefont{L.}~\bibnamefont{Kleinman}},
  \bibinfo{journal}{\textit{Phys. Rev. B}} \textbf{2004}, \textit{\bibinfo{volume}{70}},
  \bibinfo{pages}{180406-180409}.

\bibitem[{\citenamefont{Ono et~al.}(2002)\citenamefont{Ono, Okabayashi,
  Mizuguchi, Oshima, Fujimori, and Akinaga}}]{fujimori}
\bibinfo{author}{\bibfnamefont{K.}~\bibnamefont{Ono}},
  \bibinfo{author}{\bibfnamefont{J.}~\bibnamefont{Okabayashi}},
  \bibinfo{author}{\bibfnamefont{M.}~\bibnamefont{Mizuguchi}},
  \bibinfo{author}{\bibfnamefont{M.}~\bibnamefont{Oshima}},
  \bibinfo{author}{\bibfnamefont{A.}~\bibnamefont{Fujimori}}, \bibnamefont{and}
  \bibinfo{author}{\bibfnamefont{H.}~\bibnamefont{Akinaga}},
  \bibinfo{journal}{\textit{J. Appl. Phys.}} \textbf{2002}, \textit{\bibinfo{volume}{91}},
  \bibinfo{pages}{8088-8092}.

\bibitem[{\citenamefont{Kim et~al.}(2006)\citenamefont{Kim, Jeon, Kang, Lee,
  Lee, and Jin}}]{film}
\bibinfo{author}{\bibfnamefont{T.~W.} \bibnamefont{Kim}},
  \bibinfo{author}{\bibfnamefont{H.~C.} \bibnamefont{Jeon}},
  \bibinfo{author}{\bibfnamefont{T.~W.} \bibnamefont{Kang}},
  \bibinfo{author}{\bibfnamefont{H.~S.} \bibnamefont{Lee}},
  \bibinfo{author}{\bibfnamefont{J.~Y.} \bibnamefont{Lee}}, \bibnamefont{and}
  \bibinfo{author}{\bibfnamefont{S.}~\bibnamefont{Jin}},
  \bibinfo{journal}{\textit{Appl. Phys. Lett.}} \textbf{2006} \textit{\bibinfo{volume}{88}},
  \bibinfo{pages}{021915-021917}.

\bibitem[{\citenamefont{Mermin}(1979)}]{mermin}
\bibinfo{author}{\bibfnamefont{N.~D.} \bibnamefont{Mermin}},
  \bibinfo{journal}{\textit{Rev. Modern Phys.}} \textbf{1979}, textit{\bibinfo{volume}{51}},
  \bibinfo{pages}{591-648}.

\bibitem[{\citenamefont{Li and Wang}(2005)}]{biblia}
\bibinfo{author}{\bibfnamefont{J.}~\bibnamefont{Li}} \bibnamefont{and}
  \bibinfo{author}{\bibfnamefont{L.~W.} \bibnamefont{Wang}},
  \bibinfo{journal}{\textit{Phys. Rev. B}} \textbf{2005}, \textit{\bibinfo{volume}{72}},
  \bibinfo{pages}{125325-125339}.

\bibitem[{\citenamefont{Kazempour et~al.}(2009)\citenamefont{Kazempour,
  Hashemifar, and Akbarzadeh}}]{prb-79-195420}
\bibinfo{author}{\bibfnamefont{A.}~\bibnamefont{Kazempour}},
  \bibinfo{author}{\bibfnamefont{S.~J.} \bibnamefont{Hashemifar}},
  \bibnamefont{and}
  \bibinfo{author}{\bibfnamefont{H.}~\bibnamefont{Akbarzadeh}},
  \bibinfo{journal}{\textit{Phys. Rev. B}} \textbf{2009}, \textit{\bibinfo{volume}{79}},
  \bibinfo{pages}{195420-195427}.

\bibitem[{\citenamefont{Trampert et~al.}(2001)\citenamefont{Trampert, Schippan,
  D{\"a}weritz, and Ploog}}]{mismatch}
\bibinfo{author}{\bibfnamefont{A.}~\bibnamefont{Trampert}},
  \bibinfo{author}{\bibfnamefont{F.}~\bibnamefont{Schippan}},
  \bibinfo{author}{\bibfnamefont{L.}~\bibnamefont{D{\"a}weritz}},
  \bibnamefont{and} \bibinfo{author}{\bibfnamefont{H.}~\bibnamefont{Ploog}},
  \bibinfo{journal}{\textit{Appl. Phys. Lett.}} \textbf{2001}, \textit{\bibinfo{volume}{78}},
  \bibinfo{pages}{2461-2463}.

\bibitem[{\citenamefont{Di Xu et al}(2007)}]{cu}
\bibinfo{author}{\bibfnamefont{Di} \bibnamefont{Xu}},
\bibinfo{author}{\bibfnamefont{Wei Lek} \bibnamefont{Kwan}},
\bibinfo{author}{\bibfnamefont{Kai} \bibnamefont{Chen}},
\bibinfo{author}{\bibfnamefont{Xi} \bibnamefont{Zhang}},
\bibinfo{author}{\bibfnamefont{Vidvus} \bibnamefont{Ozolins}}
\bibnamefont{and}
\bibinfo{author}{\bibfnamefont{K.N. } \bibnamefont{Tu}},
\bibinfo{journal}{\textit{Appl. Phys. Lett.}} \textbf{2007},
  \textit{\bibinfo{volume}{91}}, \bibinfo{pages}{254105-254107}.


\bibitem[{\citenamefont{Goodenough and Kafalas}(1967)}]{goodenough}
\bibinfo{author}{\bibfnamefont{J.~B.} \bibnamefont{Goodenough}}
  \bibnamefont{and} \bibinfo{author}{\bibfnamefont{J.~A.}
  \bibnamefont{Kafalas}}, \bibinfo{journal}{\textit{Phys. Rev.}} \textbf{1967},
  \textit{\bibinfo{volume}{157}}, \bibinfo{pages}{389-395}.

\bibitem[{\citenamefont{Menyuk et~al.}(1969)\citenamefont{Menyuk, Kafalas,
  Dwight, and Goodenough}}]{menyuk}
\bibinfo{author}{\bibfnamefont{N.}~\bibnamefont{Menyuk}},
  \bibinfo{author}{\bibfnamefont{J.~A.} \bibnamefont{Kafalas}},
  \bibinfo{author}{\bibfnamefont{K.}~\bibnamefont{Dwight}}, \bibnamefont{and}
  \bibinfo{author}{\bibfnamefont{J.~B.} \bibnamefont{Goodenough}},
  \bibinfo{journal}{\textit{Phys. Rev.}} \textbf{1969}, \textit{\bibinfo{volume}{177}},
  \bibinfo{pages}{942-951}.

\bibitem[{\citenamefont{Continenza et~al.}(2001)\citenamefont{Continenza,
  Picozzi, Geng, and Freeman}}]{continenza}
\bibinfo{author}{\bibfnamefont{A.}~\bibnamefont{Continenza}},
  \bibinfo{author}{\bibfnamefont{S.}~\bibnamefont{Picozzi}},
  \bibinfo{author}{\bibfnamefont{W.~T.} \bibnamefont{Geng}}, \bibnamefont{and}
  \bibinfo{author}{\bibfnamefont{A.~J.} \bibnamefont{Freeman}},
  \bibinfo{journal}{\textit{Phys. Rev. B}} \textbf{2001}, \textit{\bibinfo{volume}{64}},
  \bibinfo{pages}{085204-085210}.




\bibitem[{\citenamefont{Sanvito and Hill}(2000)}]{sanvito}
\bibinfo{author}{\bibfnamefont{S.}~\bibnamefont{Sanvito}} \bibnamefont{and}
  \bibinfo{author}{\bibfnamefont{N.~A.} \bibnamefont{Hill}},
  \bibinfo{journal}{\textit{Phys. Rev. B}} \textbf{2000}, \textit{\bibinfo{volume}{62}},
  \bibinfo{pages}{15553-15560}.

\bibitem[{\citenamefont{Zhao et~al.}(2002)\citenamefont{Zhao, Geng, Freeman,
  and Delley}}]{freeman-2}
\bibinfo{author}{\bibfnamefont{Y.-J.} \bibnamefont{Zhao}},
  \bibinfo{author}{\bibfnamefont{W.~T.} \bibnamefont{Geng}},
  \bibinfo{author}{\bibfnamefont{A.~J.} \bibnamefont{Freeman}},
  \bibnamefont{and} \bibinfo{author}{\bibfnamefont{B.}~\bibnamefont{Delley}},
  \bibinfo{journal}{\textit{Phys. Rev. B}} \textbf{2002}, \textit{\bibinfo{volume}{65}},
  \bibinfo{pages}{113202-113205}.

\bibitem[{\citenamefont{Caroff et~al.}(2009)\citenamefont{Caroff, Dick,
  Johansson, Messing, Deppert, and Samuelson}}]{caroff}
\bibinfo{author}{\bibfnamefont{P.}~\bibnamefont{Caroff}},
  \bibinfo{author}{\bibfnamefont{K.~A.} \bibnamefont{Dick}},
  \bibinfo{author}{\bibfnamefont{J.}~\bibnamefont{Johansson}},
  \bibinfo{author}{\bibfnamefont{M.~E.} \bibnamefont{Messing}},
  \bibinfo{author}{\bibfnamefont{K.}~\bibnamefont{Deppert}}, \bibnamefont{and}
  \bibinfo{author}{\bibfnamefont{L.}~\bibnamefont{Samuelson}},
  \bibinfo{journal}{\textit{Nature Nanotech.}} \textbf{2009}, \textit{\bibinfo{volume}{4}},
  \bibinfo{pages}{50-55}.

\bibitem[{\citenamefont{Gudiksen et~al.}(2002)\citenamefont{Gudiksen, Lauhon,
  Wang, Smith, and Lieber}}]{gudiksen}
\bibinfo{author}{\bibfnamefont{M.~S.} \bibnamefont{Gudiksen}},
  \bibinfo{author}{\bibfnamefont{L.~J.} \bibnamefont{Lauhon}},
  \bibinfo{author}{\bibfnamefont{J.}~\bibnamefont{Wang}},
  \bibinfo{author}{\bibfnamefont{D.~C.} \bibnamefont{Smith}}, \bibnamefont{and}
  \bibinfo{author}{\bibfnamefont{C.~M.} \bibnamefont{Lieber}},
  \bibinfo{journal}{\textit{Nature}} \textbf{2002}, \textit{\bibinfo{volume}{415}},
  \bibinfo{pages}{617-620}.

\bibitem[{\citenamefont{Kresse and Hafner}(1993)}]{kresse1}
\bibinfo{author}{\bibfnamefont{G.}~\bibnamefont{Kresse}} \bibnamefont{and}
  \bibinfo{author}{\bibfnamefont{J.}~\bibnamefont{Hafner}},
  \bibinfo{journal}{\textit{Phys. Rev. B}} \textbf{1993}, \textit{\bibinfo{volume}{47}},
  \bibinfo{pages}{558-561}.

\bibitem[{\citenamefont{Kresse and Furthm{\"u}ller}(1996)}]{kresse2}
\bibinfo{author}{\bibfnamefont{G.}~\bibnamefont{Kresse}} \bibnamefont{and}
  \bibinfo{author}{\bibfnamefont{J.}~\bibnamefont{Furthm{\"u}ller}},
  \bibinfo{journal}{\textit{Phys. Rev. B}} \textbf{1996}, \textit{\bibinfo{volume}{54}},
  \bibinfo{pages}{11169-11186}.

\bibitem[{\citenamefont{Kresse and Joubert}(1999)}]{kresse3}
\bibinfo{author}{\bibfnamefont{G.}~\bibnamefont{Kresse}} \bibnamefont{and}
  \bibinfo{author}{\bibfnamefont{D.}~\bibnamefont{Joubert}},
  \bibinfo{journal}{\textit{Phys. Rev. B}} \textbf{1999}, \textit{\bibinfo{volume}{59}},
  \bibinfo{pages}{1758-1775}.

\bibitem[{\citenamefont{Kresse and Furthm{\"u}ller}(1999)}]{kresse4}
\bibinfo{author}{\bibfnamefont{G.}~\bibnamefont{Kresse}} \bibnamefont{and}
  \bibinfo{author}{\bibfnamefont{J.}~\bibnamefont{Furthm{\"u}ller}},
  \emph{\bibinfo{title}{VASP the Guide}} (\bibinfo{publisher}{Vienna University
  of Technology}, \bibinfo{address}{Viena}, \bibinfo{year}{1999}).

\bibitem[{\citenamefont{Perdew et~al.}(1996)\citenamefont{Perdew, Burke, and
  Ernzerhof}}]{pbe}
\bibinfo{author}{\bibfnamefont{J.~P.} \bibnamefont{Perdew}},
  \bibinfo{author}{\bibfnamefont{K.}~\bibnamefont{Burke}}, \bibnamefont{and}
  \bibinfo{author}{\bibfnamefont{M.}~\bibnamefont{Ernzerhof}},
  \bibinfo{journal}{\textit{Phys. Rev. Lett.}} \textbf{1996}, \textit{\bibinfo{volume}{77}},
  \bibinfo{pages}{3865-3868}.

\bibitem[{\citenamefont{Park et~al.}(2000)\citenamefont{Park, Kwon, and
  Min}}]{park}
\bibinfo{author}{\bibfnamefont{J.~H.} \bibnamefont{Park}},
  \bibinfo{author}{\bibfnamefont{S.~K.} \bibnamefont{Kwon}}, \bibnamefont{and}
  \bibinfo{author}{\bibfnamefont{B.~I.} \bibnamefont{Min}},
  \bibinfo{journal}{\textit{Physica B}} \textbf{2000}, \textit{\bibinfo{volume}{281\&282}},
  \bibinfo{pages}{703-704}.

\bibitem[{\citenamefont{K{\"a}stner et~al.}(2002)\citenamefont{K{\"a}stner,
  D{\"a}weritz, and Ploog}}]{kas}
\bibinfo{author}{\bibfnamefont{M.}~\bibnamefont{K{\"a}stner}},
  \bibinfo{author}{\bibfnamefont{L.}~\bibnamefont{D{\"a}weritz}},
  \bibnamefont{and} \bibinfo{author}{\bibfnamefont{K.~H.} \bibnamefont{Ploog}},
  \bibinfo{journal}{\textit{Surf. Sci.}} \textbf{2002}, \textit{\bibinfo{volume}{511}},
  \bibinfo{pages}{323-330}.

\bibitem[{\citenamefont{Schippan et~al.}(2000)\citenamefont{Schippan,
  K{\"a}stner, D{\"a}weritz, and Ploog}}]{schippan}
\bibinfo{author}{\bibfnamefont{F.}~\bibnamefont{Schippan}},
  \bibinfo{author}{\bibfnamefont{M.}~\bibnamefont{K{\"a}stner}},
  \bibinfo{author}{\bibfnamefont{L.}~\bibnamefont{D{\"a}weritz}},
  \bibnamefont{and} \bibinfo{author}{\bibfnamefont{K.~H.} \bibnamefont{Ploog}},
  \bibinfo{journal}{\textit{Appl. Phys. Lett.}} \textbf{2000}, \textit{\bibinfo{volume}{76}},
  \bibinfo{pages}{834-836}.

\end{thebibliography}

\end{document}